\documentclass[conference]{IEEEtran}
\usepackage[utf8]{inputenc}
\usepackage{amsmath}
\usepackage{graphicx}
\usepackage{caption}
\usepackage{epstopdf}
\usepackage{amsmath,amsfonts,amssymb,amsthm,epsfig,epstopdf,url,array}
\usepackage{graphicx}
\usepackage{color}
\usepackage{textcomp}
\newtheorem{theorem}{Theorem}

\usepackage{cite}
\usepackage{amsthm}
\usepackage{amsmath,amsthm}
\usepackage{marvosym}
\usepackage{array}
\def\BibTeX{{\rm B\kern-.05em{\sc i\kern-.025em b}\kern-.08em
    T\kern-.1667em\lower.7ex\hbox{E}\kern-.125emX}}
\begin{document}

\IEEEoverridecommandlockouts
\title{Secure HARQ-IR-Aided Terahertz Communications}
\author{
\IEEEauthorblockN{
Yongkang Li\IEEEauthorrefmark{1}, Ziyang Song\IEEEauthorrefmark{1},
Zheng Shi\textsuperscript{\Letter}\IEEEauthorrefmark{1},
Qingping Dou\IEEEauthorrefmark{1},
Hongjiang Lei\IEEEauthorrefmark{2},
Jinming Wen\IEEEauthorrefmark{1}
and Junbin Fang\IEEEauthorrefmark{1}
\thanks {This work was supported in part by National Natural Science Foundation of China under Grants 62171200 and 61971080, in part by Chongqing Key Laboratory of Mobile Communications Technology under Grant cqupt-mct-202204, in part by Guangdong Basic and Applied Basic Research Foundation under Grant 2023A1515010900, and in part by Zhuhai Basic and Applied Basic Research Foundation under Grant ZH22017003210050PWC. (\emph{Corresponding Author: Zheng Shi.})}
}
\IEEEauthorblockA{\IEEEauthorrefmark{1}School of Intelligent Systems Science and Engineering, Jinan University, Zhuhai 519070, China}
\IEEEauthorblockA{\IEEEauthorrefmark{2}Chongqing Key Lab of Mobile Communications Technology,\\
Chongqing University of Posts and Telecommunications, Chongqing 400065, China}} %
\date{December 2022}

\maketitle
\begin{abstract}
Terahertz (THz) communication is one of the most promising candidates to accommodate high-speed mobile data services. This paper proposes a secure hybrid automatic repeat request with incremental redundancy (HARQ-IR) aided THz communication scheme, where the transmission secrecy is ensured by confusing the eavesdropper with dummy messages. The connection and secrecy outage probabilities are then derived in closed-form. Besides, the tail behaviour of the connection outage probability in high signal-to-noise ratio (SNR) is examined by carrying out the asymptotic analysis, and the upper bound of the secrecy outage probability is obtained in a simple form by capitalizing on large deviations. With these results, we take a step further to investigate the secrecy long term average throughput (LTAT). By noticing that HARQ-IR not only improves the reliability of the legitimate user, but also increases the probability of being eavesdropped, a robust rate adaption policy is finally proposed to maximize the LTAT while restricting the connection and secrecy outage probabilities within satisfactory requirements.



\end{abstract}

\section{Introduction}\label{sec:int}
In order to offer higher capacity, ultra-high frequencies are foreseen to be used in the sixth generation (6G). Terahertz (THz) communications are recognized as one of key enabling technologies to deliver a peak data rate of 1 Tbps.
Unfortunately, in contrast to the low-frequency communications, THz communications suffer from severe path-loss attenuation, and are susceptible to the atmospheric turbulence, the pointing errors, the molecular absorption, etc. These negative effects degrade the reception reliability of THz communications. In \cite{8610080,9039743}, the error performance of THz communications was examined to account for the antenna misalignment and hardware imperfections. To remedy these defects, several approaches have been put forward to fulfill reliable THz communications. To be specific, an offloading strategy was proposed to minimize the energy consumption under ultra-reliable and low latency constraints in \cite{THz2_2}. In \cite{10005197}, reconfigurable reflecting surface (RIS) was leveraged to ensure the reliability and latency requirements of THz communications.
Moreover, hybrid automatic repeat request (HARQ) has been acknowledged as an effective transmission technique to boost the reliability of signal reception, albeit at the price of additional transmission delay. Thereon, the authors in \cite{THz2_3,2304.11341} thoroughly investigated the outage performance of three different HARQ types assisted THz communications.

Another inherent advantage of THz communications is the provision of physical layer security owing to its high antenna directionality \cite{9482609}, which has received considerable attentions recently. To name a few, Qiao {\it et al.} in \cite{PLS0_2} proposed a RIS-assisted secure THz transmission scheme. In \cite{9497766}, an artificial noise based mechanism was developed to address the in-beam security issue. In addition, a secure two-phase transmission strategy with unmanned aerial vehicles (UAV) relaying was devised in \cite{9709673} to safeguard THz communications. 
However, so far, there have been almost no readily available results concerned with the physical layer security of HARQ-aided THz communications in the literature. Notably, the retransmission strategy of HARQ will yield increasing probability of being eavesdropped.  This motivates us to study the performance of secure HARQ-aided THz communications from the information-theoretical perspective, with which useful system design guidelines can be extracted for THz communications. Moreover, unlike previous literature with inaccurate and complicated outage expressions, this paper provides precise expressions and deep insights into the benefits of physical layer security through approximate analysis.




In this paper, we focus on secure HARQ with incremental redundancy (HARQ-IR) aided THz communications, where the eavesdropper is confused through the introduction of dummy messages. At first, the connection and secrecy outage probabilities are derived in closed-form. With these results, the asymptotic/approximate expression of the connection/secrecy outage probability in high signal-to-noise ratio (SNR) was derived by capitalizing on the asymptotic analysis/large deviation. Besides, the secrecy long term average throughput (LTAT) is expressed in terms of the outage metrics. Furthermore, although HARQ-IR is able to improve the reliability of the legitimate user, the eavesdropping probability is increased. Therefore, we eventually develop a robust rate adaption policy to maximize the LTAT while ensuring the connection and secrecy outage constraints. 

The reminder of this paper is outlined as follows. In Section II, we introduce the system model for secure HARQ-IR aided THz communications. The outage and throughput performance metrics are then analyzed in Section III. In Section IV, the numerical results are presented for verification and a robust rate adaption policy is proposed. Section V finally concludes this paper.

\section{System Model}
\subsection{Secure HARQ-IR Transmissions}
In this paper, we consider a HARQ-IR aided THz communication system in the presence of a single passive eavesdropper. The  transmitter (i.e., Alice) sends the confidential information to the receiver (i.e., Bob) through the main channel, and the eavesdropper (i.e., Eve) overhears the transmitted signal. Herein, we assume that Alice does not know the instantaneous channel state information (CSI), which may result in the possibility of communication interruptions. In order to ensure the reception reliability of confidential message at Bob, HARQ-IR is adopted. More specifically, if the outage event occurs, Bob will request the retransmission of the message by feeding back a non-acknowledge (NACK) message. According to HARQ-IR, a new packet with different redundancy will be delivered in the next HARQ round once receiving the NACK request.

To implement the secure HARQ protocol between Alice and Bob, the Wyner codes are used \cite{4802331,8355527,6844902}. The confidential information is first encoded into a mother code of length $ML$, which is then split into $M$ sub-codewords, each with length $L$. The $M$ sub-codewords will be conveyed one by one upon request, and $M$ refers to the maximum number of transmissions. The basic idea of Wyner code is leveraging random binning approach, in which dummy message is randomly inserted into the confidential message so as to increase the secrecy level. Particularly, in order to convey the confidential message in the set ${\mathcal W} = \{1,2,\cdots,2^{LR_s}\}$, we adopt a Wyner code ${\mathcal C}(R_0/M,R_s/M,ML)$ of size $2^{LR_0}$ codewords. Therein, the two rate parameters of the Wyner code, i.e., $R_0$ and $R_s$, are the main channel code rate and the secrecy information rate, respectively. Moreover, the difference $R_0-R_s$ is termed as the secrecy gap (or called dummy message rate), which is introduced to confuse the eavesdropper. During the first HARQ round, the sub-codeword ${\bf x}_1$ is formed by a punctured Wyner code of length $L$, i.e., ${\mathcal C}(R_0,R_s,L)$. Furthermore, after $m$ HARQ rounds, all the transmitted sub-codewords constitute $[{\bf x}_1,\cdots,{\bf x}_m]$ that corresponds to a punctured code of length $mL$, i.e., ${\mathcal C}(R_0/m,R_s/m,mL)$.

By considering block fading wiretap THz channels, the received signals at the legitimate user (i.e., Bob) and eavesdropper (i.e., Eve), i.e., ${\bf y}_{B,m}$ and ${\bf y}_{E,m}$, in the $m$-th HARQ round can be expressed as
\begin{align}\label{model}
{\bf y}_{\beta,m} &= \sqrt{P_m}{h_{\beta,m}}{\bf x}_m + {{\bf n}_{\beta,m}},\,\beta\in \{B,E\},
\end{align}
where $P_m$ represents the transmit power in the $m$-th HARQ round, ${\bf{n}}_{\beta,m}$ corresponds to the complex additive white Gaussian noises (AWGN) with zero mean and variance of $N_0$, ${h_{B,m}}$ and ${h_{E,m}}$ are the THz channel coefficient of the main channel and the eavesdropper's channel, respectively.

\subsection{THz Channel Model}
 By following the THz channel modeling in \cite{9039743}, the THz channel coefficient ${h_{\beta,m}}$ can be modeled as 
 \begin{equation}\label{hm}
{{h_{\beta,m}}} = {h_{\beta,l}}{h_{\beta,pf,m}},\, \beta\in \{B,E\},
 \end{equation}
 where ${h_{\beta,l}}$ is the deterministic THz path gain and remains constant during all HARQ rounds, and ${h_{\beta,pf,m}}$ quantifies the combining influence of antenna misalignment and multipath fading. According to \cite{8610080}, ${h_{\beta,l}}$ is given by
 \begin{equation}\label{eqn:path gain}
{h_{\beta,l}} = \frac{{c\sqrt {{G_{t}}{G_{\beta}}} }}{{4\pi {f}{d_\beta}}}\exp \left( { - \frac{1}{2}\kappa ({f},T,\psi ,p){d_\beta}} \right),
\end{equation}
where $c$ and $f$ stand for the light speed and the carrier frequency, respectively, ${{G_{t}}}$ represents the transmit antenna gain, $d_\beta$ and ${{G_{\beta}}}$ are the transmission distance and receive antenna gains, respectively. 
$\kappa ({f},T,\psi ,p)$ characterizes the molecular absorption coefficient that is decided by the temperature $T$, the relative humidity $\psi$, and the atmospheric pressure $p$. As proved in \cite{8610080}, $\kappa ({f_1},T,\psi ,p)$ is explicitly obtained as
\begin{equation}
    \kappa ({f_1},T,\psi ,p) = \kappa_1(f_1,\upsilon)+\kappa_2(f_1,\upsilon)+\Lambda(f_1),
\end{equation}
where $\upsilon={\psi p_w(T,p)}/{(100p)}$,
$\upsilon$ is the volume mixing ratio of the water vapor, $p_w(T,p)$ refers to the partial pressure of saturated water vapor that depends on the temperature $T$ and pressure $p$. Besides, the terms $\kappa_1(f_1,\upsilon)$, $\kappa_2(f_1,\upsilon)$, and $\Lambda(f_1)$ can be calculated by using the simplified model of molecular absorption loss as \cite{8417891}
\begin{equation}
    \kappa_1(f_1,\upsilon) = \frac {q_1\upsilon (q_2\upsilon+q_3)} {(q_4\upsilon+q_5)^2+(\frac{f} {100c}-c_1)^2},
\end{equation}
\begin{equation}
    \kappa_2(f_1,\upsilon) = \frac {q_6\upsilon(q_7\upsilon+q_8)} {(q_9\upsilon+q_{10})^2+(\frac{f} {100c}-c_2)^2},
\end{equation}
\begin{equation}
    \Lambda(f_1) = j_1{f_1^3} + j_2{f_1^2} + j_3{f_1} + j_4,
\end{equation}
where $q_1=0.2205$, $q_2 = 0.1303$, $q_3 = 0.0294$, $q_4 = 0.4093$,
$q_5 = 0.0925$, $q_6 = 2.014$, $q_7 = 0.1702$, $q_8 = 0.0303$,
$q_9 = 0.537$, $q_{10} = 0.0956$, $c_1= 10.835$cm$^{-1}$, $c_2=12.664$cm$^{-1}$, $j_1=5.54  \times 10^{-37}$Hz$^{-3}$, $j_2=-3.94  \times 10^{-25}$Hz$^{-2}$, $j_3=9.06  \times 10^{-14}$Hz$^{-1}$, $j_4=-6.36  \times 10^{-3}$Hz$^{-3}$.
Moreover, as derived in \cite{8610080}, the probability density function (PDF) of $|{h_{\beta,pf,m}}|$ can be expressed as
\begin{equation}\label{eqn:PDF}
{f_{|{h_{\beta,pf,m}}|}}(x) = \frac{{\phi_\beta {\mu ^{\frac{\phi_\beta}{\alpha }}}{x^{\phi_\beta - 1}}}}{{S_\beta^{\phi_\beta} \hat h_{f,\beta}^{\phi_\beta} \Gamma (\mu )}}\Gamma \left( {\frac{{\alpha \mu  - \phi_\beta }}{\alpha },\frac{{\mu {x^\alpha }}}{{S_\beta^\alpha \hat h_{f,\beta}^\alpha }}} \right),
\end{equation}
where $\Gamma(a) $ and $ \Gamma \left( a,x \right)$ 
denote Gamma function and the upper incomplete Gamma function, respectively,
$\mu$ and ${\hat h_{f,\beta}}$ denote the variance and the $\alpha$-root mean value of the fading channel envelope, respectively, ${S_\beta} = {\left| {{\rm erf}(\zeta_\beta)} \right|^2}$ is the fraction of the maximum collected power over THz channels and $\zeta_\beta  = \sqrt \pi  {r_\beta}/\left( {\sqrt 2 {w_{d_\beta}}} \right)$, $r_\beta$ and ${w_{d_\beta}}$ stand for the radius of the receive antenna effective area and the transmission beam footprint radius at reference distance ${d_\beta}$, respectively, ${\phi _\beta } = w_{{d_\beta }}^2\sqrt \pi  {\rm{erf}}\left( {{\zeta _\beta }} \right)\exp (\zeta _\beta ^2)/(8{\zeta _\beta }\sigma _\beta ^2)$ and $\sigma_\beta$ are
the ratio of normalized beam-width to the jitter and the doubled spatial jitter standard deviation of THz channels, respectively.

\subsection{Achievable Mutual Information}
By following the information theory of secure HARQ-IR \cite{8355527}, the accumulated mutual information of HARQ-IR aided THz communications achieved by the legitimate user and eavesdropper after $M$ HARQ rounds can be obtained as
\begin{equation}\label{eqn:harq IR}
I_{\beta}(M) = \sum\limits_{m = 1}^M {{{\log }_2}\left(1 + {\rho _m}{{\left| {{h_{\beta,l}}} \right|}^2}{{\left| {{h_{\beta,pf,m}}} \right|}^2}\right)}, \beta\in \{B,E\},
\end{equation}
where $\rho_m=P_m/N_0$ denotes the transmit signal-to-noise ratio.
It is worth noting that the introduction of HARQ-IR is not only beneficial to enhance reception reliability for legitimate users, but also is vulnerable to eavesdropping. 
Therefore, it is imperative to examine both the connection outage and the secrecy outage. According to \cite{4802331}, the connection outage occurs if the accumulated mutual information attained by the legitimate user is below the code rate $R_0$, i.e., $I_{B}(M)< R_0$. Whereas, the secrecy outage occurs if the accumulated mutual information is larger than the dummy message rate $R_0 - R_s$, i.e., $ I_{E}(M)> R_0 - R_s$.

\section{Performance Analysis of Secure HARQ}\label{sec:opa}
 In this section, we first study the outage performance of secure HARQ-IR over THz fading channels. It has been mentioned that there are two types of outage events, i.e., the connection outage and the secrecy outage. Hence, these two types of outage probabilities are studied individually. With the analytical results, the secrecy long term average throughput (LTAT) is then evaluated.

\subsection{Connection Outage Probability}
As aforementioned, the connection outage occurs if $I_{B}(M)< R_0$. Accordingly, the connection outage probability $P_{co}$ can be obtained by averaging over all the realizations of the channel process, i.e.,
\begin{equation}\label{Pe}
  P_{co} = \Pr\left\{I_{B}(M)<R_0\right\}.
\end{equation}
In what follows, the exact analysis of $P_{co}$ is firstly performed, and the asymptotic connection outage probability is then derived in the high SNR regime, i.e., $\rho_1,\cdots,\rho_M\to \infty$.
\subsubsection{Exact Analysis}
By substituting \eqref{eqn:harq IR} into \eqref{Pe}, the derivations of the connection outage probability amount to determining the distribution of the product of multiple random variables. This inspires us to capitalize on the Mellin transform. Fortunately, as proved in \cite{2304.11341}, this method was applied to derive the distribution of the accumulated mutual information of HARQ-IR aided THz communications in closed-form, as given by the following theorem.
 \begin{theorem} \label{the:clo_c}{\cite[eq.(11)]{2304.11341}}
The cumulative distribution function (CDF) of ${I_{\beta}(M)}$ can be expressed in terms of an inverse Laplace transform as {\eqref{eqn:IR_8}}, as shown at the top of the next page, where $\rm{c} < 0$, ${\rm i} = \sqrt{-1}$, and $ H_{p,q}^{m,n}(\cdot)$ refers to the Fox’s H function {\cite{ansari2017new}}. For the notational convenience, $\Psi _\beta^M(x)$ is used to represent the CDF of ${I_{\beta}(M)}$.
 \begin{figure*}[!t]
\begin{align}\label{eqn:IR_8}
&F_{I_{\beta}(M)}(x) = \Pr\left\{I_{\beta}(M)<x\right\} \notag\\
&=\left\{ {\begin{array}{*{20}{c}}
{1 - \frac{{{\phi _\beta }{\mu ^{\frac{{{\phi _\beta }}}{\alpha }}}{x^{\frac{{{\phi _\beta }}}{2}}}}}{{\alpha S_\beta ^{{\phi _\beta }}{{\left| {{h_{\beta ,l}}} \right|}^{{\phi _\beta }}}\hat h_{f,\beta }^{{\phi _\beta }}{\rho _1}^{\frac{{{\phi _\beta }}}{2}}}}\sum\limits_{n = 0}^{\mu  - 1} {\frac{1}{{n!}}} \Gamma \left( {\frac{{\alpha n - {\phi _\beta }}}{\alpha },\frac{{\mu {x^{\frac{\alpha }{2}}}}}{{S_0^\alpha {{\left| {{h_{\beta ,l}}} \right|}^\alpha }\hat h_{f,\beta }^\alpha {\rho _1}^{\frac{\alpha }{2}}}}} \right),}&{M=1}\\
{{\left( {\frac{\phi_\beta }{{2\Gamma (\mu )}}} \right)^M}\frac{1}{{2\pi {\rm{i}}}}\int_{{{\rm{c}}} - \rm i\infty }^{{{\rm{c}}}{\rm{ + i}}\infty } {\frac{{{{e}^{ - xt\ln2}}}}{{ - t}}\prod\limits_{m = 1}^M {\frac{1}{{\Gamma \left( { - t} \right)}}H_{3,2}^{1,3}\left[ {{{\left( {\rho _m {{\left| {{h_{\beta,l}}} \right|}^2}} \right)}^{\frac{1}{2}}}{{\left( {\frac{\mu }{{\hat h_{f,\beta}^\alpha S_{\beta}^\alpha }}} \right)}^{ - \frac{1}{\alpha }}}\left| {_{\left( {0,\frac{1}{2}} \right),\left( { - \phi_\beta ,1} \right)}^{\left( {1 + t,\frac{1}{2}} \right),\left( {1 - \mu ,\frac{1}{\alpha }} \right),\left( {1 + \phi_\beta ,1} \right)}} \right.} \right]} dt},}&{M\neq1} \notag\\
\end{array}} \right.\\
&\buildrel \Delta \over = \Psi _\beta^M(x),
\end{align}
\end{figure*}
 \end{theorem}
It is noteworthy that \eqref{eqn:IR_8} can be evaluated fast and accurately by adopting the numerical inversion of Laplace transform {\cite{2304.11341}}.
Accordingly, the above theorem can be applied to express \eqref{Pe} in closed-form as
\begin{equation}\label{eqn:cn_cf}
    P_{co} = \Psi _B^M(R_0).
\end{equation}
\subsubsection{Asymptotic Analysis}
Clearly, the exact expression of the connection outage probability in \eqref{eqn:cn_cf} for $m>1$ is too complex to extract useful insights as well as ease the optimal design. Therefore, we perform the asymptotic analysis of the outage probability in the high SNR regime. 
As proved in \cite{2304.11341}, the asymptotic behaviour of (\ref{eqn:IR_8}) under the conditions of $\rho_1,\cdots,\rho_M\to \infty$ was investigated by using the residue theorem and the dominant term approximation, as given by the following theorem.
 \begin{theorem}\label{the:asy}
In the high SNR regime, the asymptotic expression of $\Psi _\beta^M(x)$ can be written as {\eqref{eqn:IR_101}} at the top of the next page, where $\theta  = \min \left\{ {\mu \alpha ,\phi_\beta } \right\}$, $\Theta = \max \left\{ {\mu \alpha ,\phi_\beta } \right\}$, ${\bf 1}_{A}(x)$ denotes the indicator function such that ${\bf 1}_{A}(x)=1$ if $x\in A$, and ${\bf 1}_{A}(x)=0$ otherwise, and $ G_{p,q}^{m,n}(\cdot)$ represents the Meijer G-function {\cite{ansari2017new}}. For the notational convenience, $\Psi _\beta^{M,\infty}(x)$ is adopted to denote the asymptotic expression of $\Psi _\beta^M(x)$ in high SNR. 
\begin{figure*}[!t]
\begin{align}\label{eqn:IR_101}
\Psi _\beta^M(x)
&\simeq 
{\left( {\frac{{\Theta {\mu ^{\frac{\theta }{\alpha } - 1}}\Gamma \left( {\frac{\theta }{2} + 1} \right)\Gamma \left( {\frac{{\alpha \mu  - \theta }}{\alpha } + 1} \right)}}{{{{\left( { - 1} \right)}^{{\rm{sgn}}\left( {\mu \alpha  - \phi_\beta } \right)}}\left( {\alpha \mu  - \phi_\beta } \right){{\left| {{h_{\beta,l}}} \right|}^\theta }\hat h_{f,\beta}^\theta S_{\beta}^\theta \Gamma (\mu )}}} \right)^M}G_{M,M}^{0,M}\left( {{2^{{x}}}\left| {\begin{array}{*{20}{c}}
{1 + \frac{\theta }{2},1 + \frac{\theta }{2},...,1 + \frac{\theta }{2}}\\
{0,1,...,1}
\end{array}} \right.} \right){\left( {\prod\limits_{m = 1}^M {{\rho_m}} } \right)^{ - \frac{\theta }{2}}}\notag\\
&\triangleq \Psi _\beta^{M,\infty}(x),
\end{align}
\end{figure*}
 \end{theorem}
With the asymptotic result in Theorem \ref{the:asy}, the connection outage probability $P_{co}$ is asymptotic to
\begin{equation}\label{eqn:asy}
    P_{co} \simeq \Psi _B^{M,\infty}(R_0).
\end{equation}

To conserve space, the in-depth discussions of the asymptotic expression \eqref{eqn:asy} are omitted here and interested readers are referred to \cite{2304.11341} for more details.

\subsection{Secrecy Outage Probability}
According to \cite{4802331}, by using the law of total probability, we can get the secrecy outage probability $P_{so}$ as
 \begin{equation}\label{Ps}
  P_{so} = \sum\limits_{m = 1}^M{\Pr\left\{\mathcal{M}=m\right\}\Pr\left\{{I_{E}(m)>R_0 - R_s}\right\}},
\end{equation}
where ${\mathcal{M}}$ denotes the number of transmissions within one HARQ cycle, that is, the number of HARQ rounds required to convey a single information message. In what follows, the exact and the approximate expressions of $P_{so}$ are derived. 
\subsubsection{Exact Analysis}
Clearly, since the distribution of $\mathcal{M}$ depends on whether Bob successfully receives the message over the main channels or the maximum number of transmissions is reached, and the probability mass function (pmf) of $\mathcal{M}$ can be obtained as \eqref{PDF_M},
\begin{figure*}[!t]
\begin{equation}\label{PDF_M}
\Pr \left\{ {{\cal M} = m} \right\} = \left\{ {\begin{array}{*{20}{c}}
{1 - \Pr \left\{ {{I_{B}}(m) < {R_0}} \right\},}&{m = 1}\\
{\Pr \left\{ {{I_{B}}(m - 1) < {R_0}} \right\} - \Pr \left\{ {{I_{B}}(m) < {R_0}} \right\},}&{m = 2,...,M - 1}\\
{\Pr \left\{ {{I_{B}}(M - 1) < {R_0}} \right\},}&{m = M}
\end{array}} \right.,
\end{equation}
\end{figure*}
as shown at the top of the next page. By applying Theorem \ref{the:clo_c} to \eqref{PDF_M}, the pmf of $\mathcal{M}$ can be evaluated.

Moreover, with regard to the term $\Pr\left\{I_{E}(M)> R_0 - R_s\right\}$ in \eqref{Ps}, note that this complementary CDF (CCDF) follows as $\Pr\left\{I_{E}(M) > R_0 - R_s\right\} = 1-\Pr\left\{I_{E}(M) < R_0 - R_s\right\}$. Hence, $\Pr\left\{I_{E}(M) > R_0 - R_s\right\}$ can also be obtained by using Theorem \ref{the:clo_c} as
\begin{equation}\label{Pr}
\begin{aligned}
     \Pr\left\{I_{E}(M)> R_0 - R_s\right\} &= 1 - \Psi _E^M(R_0-R_s).
\end{aligned}
\end{equation}
By substituting (\ref{Pr}) into (\ref{Ps}), $P_{so}$ can be consequently expressed as
\begin{equation}
    \begin{aligned}\label{PSO_1}
        P_{so} =& \sum\limits_{m = 1}^{M-1}(\Psi_B^{m-1}(R_0)-\Psi_B^{m}(R_0))
        (1-\Psi_E^m(R_0-R_s))\\
        &+\Psi_B^{M-1}(R_0)(1-\Psi_E^{M}(R_0-R_s)),
    \end{aligned}
\end{equation}
where we stipulate $\Psi _B^0(R_0)=1$.
\subsubsection{Approximate Analysis}
It is obvious that the secrecy outage probability in \eqref{PSO_1} is too cumbersome to facilitate the system design. In the meantime, it is different from the analysis of $P_{co}$ that the investigations into the asymptotic expression of $P_{so}$ in high SNR are meaningless. This is because $P_{so}$ tends to 1 under high SNR (i.e., $\rho_1,\cdots,\rho_M\to \infty$) if we directly apply Theorem \ref{the:clo_c} to \eqref{PSO_1}. In other words, the confidential message will be almost surely intercepted by eavesdropper in the high SNR regime. To overcome the shortcoming of the asymptotic analysis, an approximate result of the secrecy outage probability can be obtained based on the large deviation \cite{400650}. By assuming uniform power allocation, i.e., $\rho_1=\cdots=\rho_M$, the CCDF of ${I_{E}}(M)$ is upper bounded by using the large deviation as
\begin{align}\label{large}
\Pr\left\{{I_{E}}(M) > {R_0} - {R_s}\right\} &= \Pr\left(\sum\limits_{m = 1}^M { Z_m  > Mr_M}\right) \notag\\
&\le {e^{ - M\mathcal{I}(r_M)}},
\end{align}
where $r_M = (R_0-R_s)/M$, $Z_m={\log_2 (1 + \rho_m|{h_{E,l}}{|^2}|{h_{E,pf,m}}{|^2})}$, $\mathcal{I}(r_M) = \mathop {\max }\nolimits_{s \ge 0} \{ sr_M - \lambda (s)\}$ denotes the rate function,
$\lambda (s) = \ln \mathbb E[{e^{  s{Z_m}}}]$ is the logarithmic moment generating function (MGF) of $Z_m$, and $\mathbb E$ refers to the expectation operator.
As proved in Appendix \ref{eqn:8}, the MGF of $Z_m$ can be derived as
{\begin{multline}\label{E}
    \mathbb E[{e^{  s{Z_m}}}] =
    \frac{{\xi {{\rho _m}^{ - \frac{\phi_E }{2}}{{\left| {{h_{E,l}}} \right|}}^{ - {\phi_E }}}}}{{\Gamma ({s/\ln2})}}\\
    \times H_{2,3}^{3,1}\left[{\frac{\mathcal{C}}{{{{\rho _m}^{\frac{\alpha }{2}}{{\left| {{h_{E,l}}} \right|}}^{{\alpha }}}}}\left|{_{(0,1),(K,1),( - \frac{\phi_E }{2} - {s/\ln2},\frac{\alpha }{2})}^{(1 - \frac{\phi_E }{2},\frac{\alpha }{2}),(1,1)}}\right.}\right].
\end{multline}
where
$\mathcal{C} ={u}{{\left({S_E}{\hat h_{f,E}}\right)^{-\alpha} }}$,
 $\xi  ={{\phi_E {u^{\frac{{ \phi_E }}{\alpha }}}}} {{{S_E}^{-\phi_E} {\hat h_{f,E}}^{-\phi_E} {\Gamma (u)^{-1}}}}$, $K ={ u - \alpha^{-1}\phi_E } $.  }
According to the theory of large deviation, for a sufficiently large $M$, the secrecy outage probability approaches to the upper bound if $r_M < \mathbb E[Z_m]$. This justifies the significance of the upper bound given by \eqref{large}. 
The same approach also applies to derive the upper bound of the CCDF of ${I_{E}}(m)$ for $m<M$. By considering the upper bound of the CCDF of ${I_{E}}(m)$ for $m>1$, the secrecy outage probability of \eqref{PSO_1} is upper bounded as
\begin{align}\label{eqn:up_so}
        {P_{so}} \le& \left( {1 - \Psi _B^1\left( {{R_0}} \right)} \right)\left( {1 - \Psi _E^1\left( {{R_0} - {R_s}} \right)} \right)  \notag\\
        &+ \sum\limits_{m = 2}^{M - 1} {\left( {\Psi _B^{m - 1}\left( {{R_0}} \right) - \Psi _B^m\left( {{R_0}} \right)} \right){e^{ - m{\cal I}({r_m})}}}  \notag\\
        &+ \Psi _B^{M - 1}\left( {{R_0}} \right){e^{ - M{\cal I}({r_M})}},
\end{align}
where $r_m = (R_0-R_s)/m$. To further reduce the computational complexity, by substituting the asymptotic result of $ {\Psi _B^{m }\left( {{R_0}} \right)}$ in \eqref{eqn:asy} for $m>1$ into \eqref{eqn:up_so}, the secrecy outage probability can be approximated as
\begin{align}\label{eqn:se_appr}
    {P_{so}} \approx &
\left( {1 - \Psi _B^{1}\left( {{R_0}} \right)} \right)\left( {1 - \Psi _E^{1}\left( {{R_0} - {R_s}} \right)} \right) \notag\\
 & + \sum\limits_{m = 2}^{M - 1} {\left( {\Psi _B^{m - 1,\infty }\left( {{R_0}} \right) - \Psi _B^{m,\infty }\left( {{R_0}} \right)} \right){e^{ - m{\cal I}({r_m})}}} \notag\\
 & + \Psi _B^{M - 1,\infty }\left( {{R_0}} \right){e^{ - M{\cal I}({r_M})}}.
\end{align}

\subsection{Secrecy LTAT}\label{sec:asy_I}
According to {\cite{4802331}}, the secrecy long term average throughput (LTAT) can be expressed as
\begin{equation}\label{sec}
\begin{aligned}
  \eta = 
  \frac{R_s(1-\Psi _B^M\left( {{R_0}} \right))}{\mathbb E[\mathcal{M}]},
\end{aligned}
\end{equation}
where $\mathbb E[\mathcal{M}]$ indicates the expected number of transmissions and is given by
\begin{equation}
    \mathbb E[\mathcal{M}] = \sum\limits_{m=1}^{M}m \Pr\left\{\mathcal{M}=m\right\} = 1+\sum\limits_{m=1}^{M-1}\Psi _B^m\left( {{R_0}} \right).
\end{equation}
Furthermore, the asymptotic expression of the secure LTAT can be calculated by replacing $\Psi _B^m\left( {{R_0}} \right)$ in \eqref{sec} with $\Psi _B^{M,\infty}(R_0)$. It can be proved that the asymptotic expression of $\eta$ actually offers a lower bound of the actual secure LTAT, and interested readers are referred to \cite{7959548} for the details.


\section{Numerical Results}\label{sec:NR}
In this section, Monte Carlo simulations are conducted to verify our analytical results. Unless otherwise noted, the system parameters are set as follows, $\alpha  = 1$, $\mu  = 2$, $\psi  = 50\% $, $T = 296{\rm{^\circ }}$~K, $p = 101325$~Pa, ${\sigma _{\rm{s}}} = 1$, and the carrier frequency is $f=275$~GHz. Moreover, both the transmit and the Bob's receive antenna gains are equal to $55$~dBi, the Eve's receive antenna gain is $50$~dBi, the transmission distances from Alice to Bob and Eve are assumed to be $20$~m and $40$~m, respectively. Besides, the main code rate and the confidential rate are set as $R_0$ = 3 bps/Hz and $R_s$ = 2 bps/Hz, respectively, and the transmit SNRs during all HARQ rounds are assumed to be the same, i.e., $\rho_1=\cdots=\rho_M\triangleq \gamma_T$.

\subsection{Performance Evaluation}
\begin{figure}[htp]
    \centering
    \includegraphics[width=1.9in]{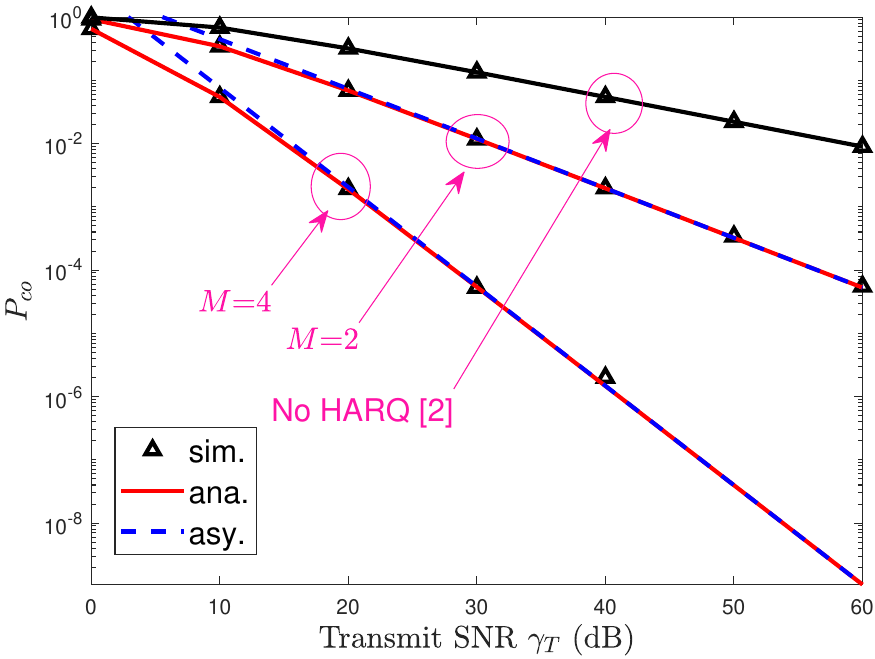}
    \caption{Connection outage probability $P_{co}$ versus $\gamma_T$.}
    \label{fig:1}
\end{figure}


\begin{figure}[htp]
    \centering
    \includegraphics[width=1.9in]{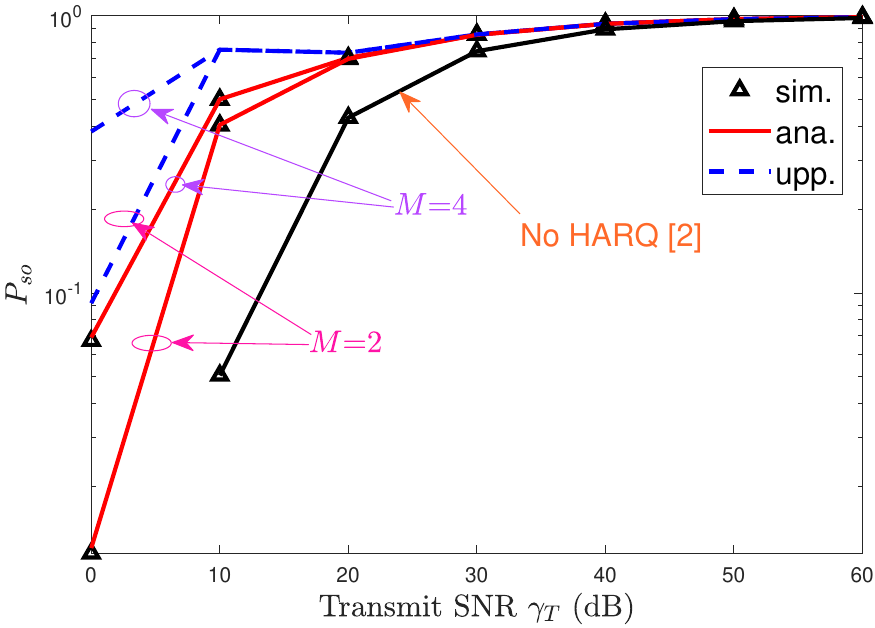}
    \caption{Secrecy outage probability $P_{so}$ versus $\gamma_T$.}
    \label{fig:2}
\end{figure}


Figs. \ref{fig:1} and \ref{fig:2} depict the connection outage probability $P_{co}$ and the secrecy outage probability $P_{so}$ 
versus the average transmit SNR $\gamma_T$, respectively. It is clearly seen from both figures that the exact and simulated results are in perfect agreement. One can also observe from Fig. \ref{fig:1} that the asymptotic results tightly approach to the exact ones with the increase of the SNR. Furthermore, the upper bound of the secrecy outage probability (labeled as ``upp.'') in Fig. \ref{fig:2} is plotted according to \eqref{eqn:up_so}, which justifies the validity of our analysis. Moreover, it can be observed that the proposed HARQ-aided communications can significantly reduce the connection outage probability, albeit at the cost of increasing the secrecy outage probability by comparing to THz communications without HARQ  (labeled as ``No HARQ'') \cite{9039743}. Hence, it is necessary to properly devise the HARQ-aided scheme to guarantee both the secrecy and the connection outage requirements. More discussions are deferred to the next subsection.

\begin{figure}[htp]
    \centering
    \includegraphics[width=1.9in]{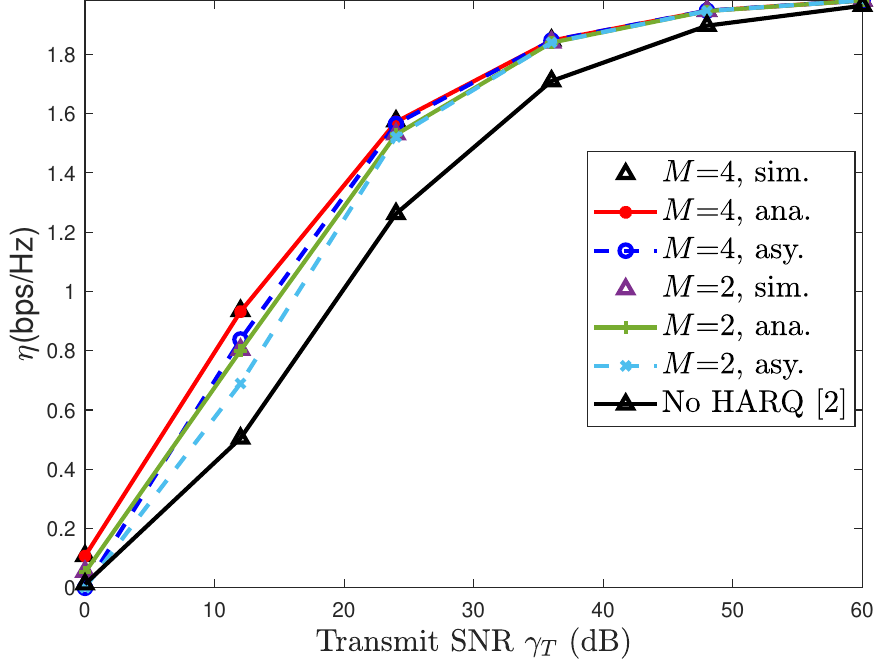}
    \caption{Secrecy LTAT $\eta$ versus $\gamma_T$.}
    \label{fig:3}
\end{figure}

Fig. \ref{fig:3} presents the secrecy LTAT $\eta$ versus the average transmit SNR. It can be seen from Fig. \ref{fig:3} that the analytical results coincide with the simulated ones. 
As expected, the asymptotic results of the secrecy LTAT in Fig. \ref{fig:3} indeed act as a lower bound of $\eta$, which is consistent with the theoretical analysis in \cite{7959548}. Furthermore, it can be observed that the proposed HARQ-IR-aided THz communications outperform THz communications without HARQ in terms of LTAT.

\subsection{Robust Design of Rate Adaption}
It should be noticed that HARQ-IR not only enhances the reliability of the legitimate user, but also increases the probability of being eavesdropped. To address this issue, the main code rate $R_0$ and confidential information rate $R_s$ are optimally designed to maximize the secure LTAT meanwhile guaranteeing the connection and the secrecy outage probabilities. The problem can be mathematically formulated as
\begin{equation}\label{eqn:rate}
\begin{array}{*{20}{c}}
{\mathop {\max }\limits_{R_0,R_s} }&{{\eta}}\\
{{\rm{s}}{\rm{.t.}}}&P_{ co}\leq\varepsilon_c,\\
                    &P_{so}\leq\varepsilon_e
\end{array}
\end{equation}
where $\varepsilon_c$ and $\varepsilon_e$ represent the maximum allowable connection outage probability and secrecy outage probability, respectively. 
In order to reduce the computational complexity, the asymptotic connection outage probability in \eqref{eqn:asy} and the approximate secrecy outage probability in \eqref{eqn:se_appr} are used to replace the exact expressions in \eqref{eqn:rate}. Since both the asymptotic results and the approximate results can be regarded as the worst-case performance benchmark, the proposed rate adaption method is actually a robust design for secure HARQ-IR-aided THz communications. 

\begin{figure}[htp]
    \centering
    \includegraphics[width=1.9in]{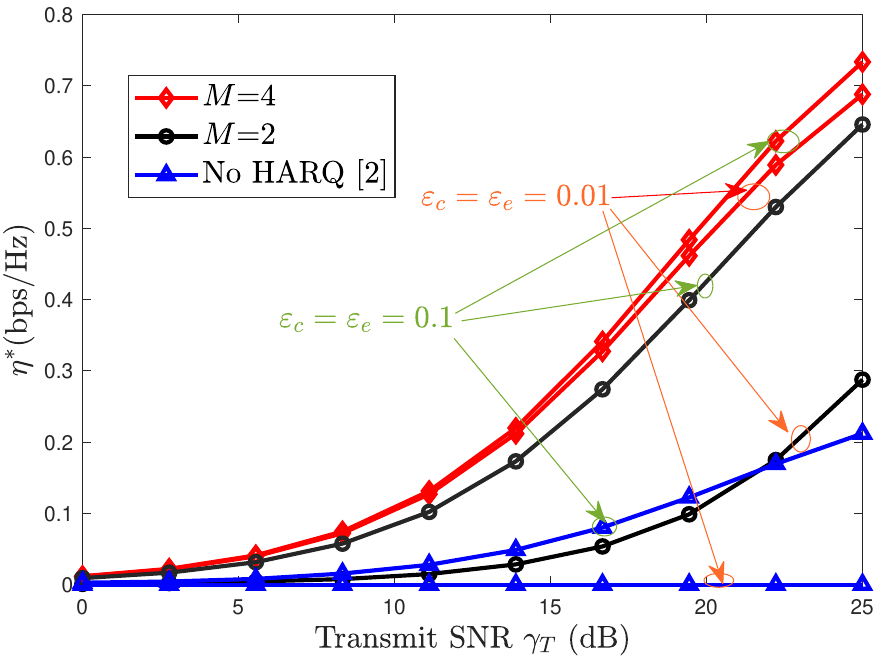}
    \caption{The optimal LTAT $\eta^*$ versus $\gamma_T$ under the condition of $\varepsilon_c=\varepsilon_e=0.01$ and $\varepsilon_c=\varepsilon_e=0.1$.}
    \label{fig:4}
\end{figure}

Fig. \ref{fig:4} illustrates the optimal secrecy LTAT versus the the average transmit SNR under different $M$. It is also observed that the increase of the maximum number of transmission is favorable for the improvement of the secure LTAT. In particular, it is clearly observed that the proposed HARQ-IR-aided scheme achieves a considerable LTAT improvement by comparing to No-HARQ, especially under a strict maximum allowable outage tolerance. 

\section{Conclusions}\label{sec:con}
In this paper, we confined our analysis to the outage and throughput performance of secure HARQ-IR assisted THz communications. More specifically, the connection outage probability was derived in closed-form, with which its asymptotic analysis in high SNR was performed. Then the secrecy outage performance was examined by conducting exact and approximate analyses. With these basis results, the secrecy LTAT was calculated accordingly. Finally, a robust rate adaption policy was proposed to maximize the LTAT while guaranteeing the outage constraints.

\appendices

\section{Proof of \eqref{E}}\label{eqn:8}
The MGF of $Z_m$, i.e., $\mathbb E[{e^{  s{Z_i}}}]$, can be written as
\begin{equation}\label{eqn:mgf_deff1}
    \mathbb E[{e^{  s{Z_i}}}]    = \mathbb E[{(1 + {\rho _m}{{\left| {{h_{E,l}}} \right|}^2}|{g_{f,{x_i}}}{|^2})^{  {s/\ln2}}}].
\end{equation}
By putting the PDF of \eqref{eqn:PDF} into \eqref{eqn:mgf_deff1}, it follows that
\begin{equation}\label{eqn:mgf_deff2}
   \mathbb E[{e^{  s{Z_i}}}] = \xi \int_0^\infty  ({1 + } {\rho _m}{{\left| {{h_{E,l}}} \right|}^2}{x^2}){^{ - {s/\ln2}}}{x^{\phi_E  - 1}}\Gamma (K,\mathcal{C}{x^\alpha })dx.
\end{equation}
Moreover, by employing the representations of the Meijer G-functions \cite[eq.(8.4.2.5), eq.(8.4.16.2)]{brychkov1986integrals}, \eqref{eqn:mgf_deff2} can be rewritten as
\begin{align}\label{z}
    & \mathbb E[{e^{  s{Z_i}}}] = \frac{\xi }{{\Gamma ({s/\ln2})}} \times\notag\\
    &  \int_0^\infty  {{x^{\phi_E  - 1}}G_{1,1}^{1,1}\left[{\rho _m}{{\left| {{h_{E,l}}} \right|}^2}{x^2}\left|{_0^{ - {s/\ln2} + 1}}\right.\right]G_{1,2}^{2,0}\left[\mathcal{C}{x^\alpha }\left|{_{0,K}^1}\right.\right]dx}.
\end{align}
By change of variable $z = x^2$ and identifies this integral with the representation of Fox's H function {\cite[eq.(21)]{adamchik1990algorithm} and \cite [eq.(8.3.2.21)]{brychkov1986integrals}}, (\ref{z}) can be consequently derived as \eqref{E}.

	\bibliographystyle{IEEEtran}
 	\bibliography{ref} 

\begin{thebibliography}{10}
\providecommand{\url}[1]{#1}
\csname url@samestyle\endcsname
\providecommand{\newblock}{\relax}
\providecommand{\bibinfo}[2]{#2}
\providecommand{\BIBentrySTDinterwordspacing}{\spaceskip=0pt\relax}
\providecommand{\BIBentryALTinterwordstretchfactor}{4}
\providecommand{\BIBentryALTinterwordspacing}{\spaceskip=\fontdimen2\font plus
\BIBentryALTinterwordstretchfactor\fontdimen3\font minus
  \fontdimen4\font\relax}
\providecommand{\BIBforeignlanguage}[2]{{%
\expandafter\ifx\csname l@#1\endcsname\relax
\typeout{** WARNING: IEEEtran.bst: No hyphenation pattern has been}%
\typeout{** loaded for the language `#1'. Using the pattern for}%
\typeout{** the default language instead.}%
\else
\language=\csname l@#1\endcsname
\fi
#2}}
\providecommand{\BIBdecl}{\relax}
\BIBdecl

\bibitem{8610080}
A.-A.~A. Boulogeorgos and A.~Alexiou, ``Error analysis of mixed {TH}z-{RF}
  wireless systems,'' \emph{IEEE Commun. Lett.}, vol.~24, no.~2, pp. 277--281,
  Feb. 2020.

\bibitem{9039743}
E.~N. Papasotiriou, A.-A.~A. Boulogeorgos, and A.~Alexiou, ``Performance
  analysis of {TH}z wireless systems in the presence of antenna misalignment
  and phase noise,'' \emph{IEEE Commun. Lett.}, vol.~24, no.~6, pp. 1211--1215,
  Jun. 2020.

\bibitem{THz2_2}
S.~Xie, H.~Li, L.~Li, Z.~Chen, and S.~Li, ``Reliable and energy-aware job
  offloading at terahertz frequencies for mobile edge computing,'' \emph{China
  Commun.}, vol.~17, no.~12, pp. 17--36, Dec. 2020.

\bibitem{10005197}
H.~Zarini, N.~Gholipoor, M.~R. Mili, M.~Rasti, H.~Tabassum, and E.~Hossain,
  ``Resource management for multiplexing e{MBB} and {URLLC} services over
  {RIS}-{A}ided {TH}z communication,'' \emph{{IEEE} Trans. Commun.}, vol.~71,
  no.~2, pp. 1207--1225, Jan. 2023.

\bibitem{THz2_3}
Z.~Song, J.~Feng, Z.~Shi, Q.~Dou, G.~Yang, Y.~Li, and S.~Ma, ``Outage
  probability analysis of {HARQ}-{A}ided terahertz communications,'' in
  \emph{Proc. Int. Conf. Wireless Commun. and Signal Process. (WCSP)}.\hskip
  1em plus 0.5em minus 0.4em\relax IEEE, 2021, pp. 1--6.

\bibitem{2304.11341}
Z.~Song, Z.~Shi, J.~Su, Q.~Dou, G.~Yang, H.~Ding, and S.~Ma, ``Performance
  analysis and optimal design of {HARQ}-{IR}-{A}ided terahertz
  communications,'' accepted by IEEE Trans. Veh. Technol., May 2023.

\bibitem{9482609}
P.~Porambage, G.~Gür, D.~P. Moya~Osorio, M.~Livanage, and M.~Ylianttila,
  ``6{G} security challenges and potential solutions,'' in \emph{Proc. IEEE
  Joint Eur. Conf. Netw. Commun. (EuCNC) 6G Summit}, 2021, pp. 622--627.

\bibitem{PLS0_2}
J.~Qiao and M.-S. Alouini, ``Secure transmission for intelligent reflecting
  surface-assisted mm{W}ave and terahertz systems,'' \emph{IEEE Wireless
  Commun. Lett.}, vol.~9, no.~10, pp. 1743--1747, Oct. 2020.

\bibitem{9497766}
W.~Gao, C.~Han, and Z.~Chen, ``{DNN}-{P}owered {SIC}-free receiver artificial
  noise aided terahertz secure communications with randomly distributed
  eavesdroppers,'' \emph{IEEE Trans. Wirel. Commun.}, vol.~21, no.~1, pp.
  563--576, Jul. 2021.

\bibitem{9709673}
M.~T. Mamaghani and Y.~Hong, ``Terahertz meets untrusted {UAV}-relaying:
  Minimum secrecy energy efficiency maximization via trajectory and
  communication co-design,'' \emph{IEEE Trans. Veh. Technol.}, vol.~71, no.~5,
  pp. 4991--5006, Feb. 2022.

\bibitem{4802331}
X.~Tang, R.~Liu, P.~Spasojevic, and H.~V. Poor, ``On the throughput of secure
  hybrid-{ARQ} protocols for {G}aussian block-fading channels,'' \emph{IEEE
  Trans. Inf. Theory}, vol.~55, no.~4, pp. 1575--1591, Apr. 2009.

\bibitem{8355527}
M.~Le~Treust, L.~Szczecinski, and F.~Labeau, ``Rate adaptation for secure
  {HARQ} protocols,'' \emph{IEEE Trans. Inf. Forensics Security}, vol.~13,
  no.~12, pp. 2981--2994, Dec. 2018.

\bibitem{6844902}
Y.~Sarikaya, O.~Ercetin, and C.~E. Koksal, ``Confidentiality-preserving control
  of uplink cellular wireless networks using hybrid {ARQ},'' \emph{IEEE/ACM
  Trans. Netw.}, vol.~23, no.~5, pp. 1457--1470, Oct. 2015.

\bibitem{8417891}
A.-A.~A. Boulogeorgos, E.~N. Papasotiriou, J.~Kokkoniemi, J.~Lehtomaeki,
  A.~Alexiou, and M.~Juntti, ``Performance evaluation of {TH}z wireless systems
  operating in 275-400 {GH}z band,'' in \emph{Proc. IEEE Veh. Technol. Conf},
  2018, pp. 1--5.

\bibitem{ansari2017new}
I.~S. Ansari, F.~Yilmaz, M.-S. Alouini, and O.~Kucur, ``New results on the sum
  of {G}amma random variates with application to the performance of wireless
  communication systems over {N}akagami-m fading channels,'' \emph{Wiley Trans.
  Emerging Technol. Telecommun.}, vol.~28, no.~1, p. e2912, Dec. 2014.

\bibitem{400650}
A.~Weiss, ``An introduction to large deviations for communication networks,''
  \emph{IEEE J. Sel. Areas Commun.}, vol.~13, no.~6, pp. 938--952, Aug. 1995.

\bibitem{7959548}
Z.~Shi, S.~Ma, G.~Yang, K.-W. Tam, and M.~Xia, ``Asymptotic outage analysis of
  {HARQ-IR} over time-correlated {Nakagami-$m$} fading channels,'' \emph{IEEE
  Trans. Wireless Commun.}, vol.~16, no.~9, pp. 6119--6134, Jun. 2017.

\bibitem{brychkov1986integrals}
Y.~A. Brychkov, O.~Marichev, and A.~Prudnikov, ``Integrals and {S}eries, vol 3:
  more special functions,'' \emph{New York, NY, USA: Gordon and Breach Science
  Publishers}, 1986.

\bibitem{adamchik1990algorithm}
V.~S. Adamchik and O.~Marichev, ``The algorithm for calculating integrals of
  hypergeometric type functions and its realization in reduce system,'' in
  \emph{Proc. Int. Symp. Symbolic Algebr . Comput. (ISSAC), Tokyo, Japan}, Jul.
  1990, pp. 212--224.

\end{thebibliography}
	



\end{document}